\PassOptionsToPackage{table}{xcolor}
\documentclass[sigconf,natbib=True]{acmart}

\usepackage{multirow}
\usepackage{cleveref}

\Crefname{equation}{Eq.}{Eqs.}
\Crefname{figure}{Fig.}{Figs.}
\Crefname{table}{Tab.}{Tabs.} 
\Crefname{appendix}{App.}{Apps.} 
\crefformat{section}{\S#2#1#3}

\newcommand{\eff}{\textsc{Eff}}
\newcommand{\fair}{\textsc{Fair}}
\newcommand{\ind}{$_\text{ind}$}


\copyrightyear{2025}
\acmYear{2025}
\setcopyright{cc}
\setcctype{by-sa}
\acmConference[RecSys '25]{Proceedings of the Nineteenth ACM Conference on
Recommender Systems}{September 22--26, 2025}{Prague, Czech Republic}
\acmBooktitle{Proceedings of the Nineteenth ACM Conference on Recommender
Systems (RecSys '25), September 22--26, 2025, Prague, Czech
Republic}
\acmDOI{10.1145/3705328.3748031}
\acmISBN{979-8-4007-1364-4/2025/09}

\begin{document}

\title{Stairway to Fairness: Connecting Group and Individual Fairness}

\author{Theresia Veronika Rampisela}
\authornote{Part of this work was done while visiting ADM+S at RMIT University.}
\orcid{0000-0003-1233-7690}
\affiliation{%
  \institution{University of Copenhagen}
  \city{Copenhagen}
  \country{Denmark}}
\email{thra@di.ku.dk}

\author{Maria Maistro}
\orcid{0000-0002-7001-4817}
\affiliation{%
 \institution{University of Copenhagen}
 \city{Copenhagen}
 \country{Denmark}
}
\email{mm@di.ku.dk}

\author{Tuukka Ruotsalo}
\orcid{0000-0002-2203-4928}
\affiliation{%
 \institution{University of Copenhagen}
 \city{Copenhagen}
 \country{Denmark}
}
\affiliation{%
 \institution{LUT University}
 \city{Lahti}
 \country{Finland}
}
\email{tr@di.ku.dk}

\author{Falk Scholer}
\orcid{0000-0001-9094-0810}
\affiliation{
    \institution{RMIT University}
    \city{Melbourne}
    \country{Australia}
    }
\email{falk.scholer@rmit.edu.au}

\author{Christina Lioma}
\orcid{0000-0003-2600-2701}
\affiliation{%
 \institution{University of Copenhagen}
 \city{Copenhagen}
 \country{Denmark}
}
\email{c.lioma@di.ku.dk}

\renewcommand{\shortauthors}{Rampisela et al.}

\begin{abstract}
Fairness in recommender systems (RSs) is commonly categorised into group fairness and individual fairness. However, there is no established scientific understanding of the relationship between the two fairness types, as prior work on both types has used different evaluation measures or evaluation objectives for each fairness type, thereby not allowing for a proper comparison of the two. As a result, it is currently not known how increasing one type of fairness may affect the other. To fill this gap, we study the relationship of group and individual fairness through a comprehensive comparison of evaluation measures that can be used for both fairness types. 
Our experiments with 8 runs across 3 datasets show that recommendations that are highly fair for groups 
can be very unfair for individuals. 
Our finding is novel and useful for RS practitioners aiming to improve the fairness of their systems. Our code is available at: \url{https://github.com/theresiavr/stairway-to-fairness}.
\end{abstract}



\maketitle

\section{Introduction}

With recent legislations that mandate responsible artificial intelligence development, Recommender System (RS) fairness evaluation has become increasingly important to ensure that users are not systematically disadvantaged. 
Fairness in RSs can be evaluated for groups and for individuals. \textit{Group fairness} typically refers to having equitable outcome across groups (e.g., similar effectiveness between groups of users \cite{Ekstrand2018AllAzpiazu}), while \textit{individual fairness} is commonly defined as treating similar users/items equally \cite{Wang2022} (e.g., similar effectiveness for all users \cite{Wu2021TFROM:Providers}).  
Conceptually, prior work \cite{Li2023FairnessApplications,Do2023FairnessThesis,Singh2018FairnessRankings} discusses how RSs can be fair to groups and at the same time unfair to individuals, or vice versa, but no work has empirically studied how this practically occurs in RS fairness evaluation. 

\begin{figure}
   \includegraphics[width=0.75\columnwidth, trim=1.1cm 0 0 0]{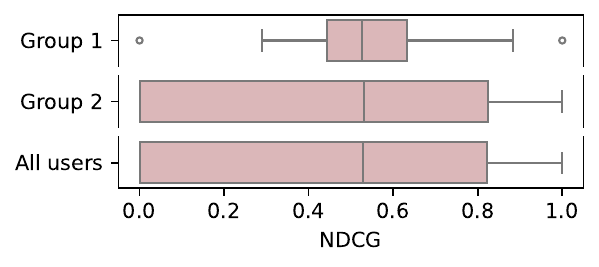}
   \centering
    \resizebox{0.55\columnwidth}{!}{
   \begin{tabular}{lccc}
   \toprule
      & \#user & $\uparrow$NDCG & $\downarrow$Gini \\
   \midrule
   \midrule
   Group 1  & 11 & 0.546 & \multirow{2}{*}{{\LARGE\}}\raisebox{1pt}{0.037}} \\ 
   Group 2  & 901 & 0.471& \\
   \midrule
   \midrule
   Individual & 912 & 0.472 & 0.446  \\ 
   \bottomrule
   \end{tabular}}

   \caption{Top: NDCG score distribution of two handpicked, non-overlapping user groups from our experiment and for all individual users of the two groups. Circles denote outliers. Bottom: Fairness score (Gini) for groups and individuals. The lower the Gini score, the fairer. Fairness is better between groups than across all individuals as both groups have similar average NDCG, but within-group variance is high, which means that recommendation quality varies widely across users (see the boxplots).
    }
    \label{fig:teaser}
\end{figure}

Prior work either: 
(i) evaluates fairness exclusively for groups or individuals \cite{Deldjoo2024FairnessDirections}; or 
(ii) evaluates both, but with two different families of measures \cite{Ferraro2024GenderImbalance,Pastor2024IntersectionalDivergence,pellegrini2023fairnessallinvestigatingharms} or for two fairness subjects/objectives \cite{Rastegarpanah2019FightingSystems, Wu2021TFROM:Providers}. 
Evaluating group and individual fairness with different families of measures makes comparison difficult, as the measure scores may differ in sensitivity, or in theoretical and empirical ranges \cite{Rampisela2024CanRelevance,Rampisela2024EvaluationStudy,Schumacher2024PropertiesRankings}. 
Likewise, it is not possible to properly compare group and individual fairness when each is evaluated for a distinct fairness objective, e.g., recommendation effectiveness disparity across individual users vs.~exposure disparity between item groups \cite{Wu2021TFROM:Providers}. To address this gap, we evaluate user-side group and individual fairness with the same families of measures that can quantify both. An example of such measures is the Gini Index (Gini) \cite{Gini1912VariabilitaMutabilita}. In \Cref{fig:teaser}, we evaluate group and individual fairness with Gini on real data and exemplify how an RS can be very fair towards user groups and at the same time much more unfair towards individual users.

In this paper, we study the relationship between evaluation measures of user-side group and individual fairness. This work is the first empirical study that compares the 9 existing user-side fairness evaluation measures for groups with those for individuals. We ask the following research questions (RQs):

\begin{enumerate}
     \item[\textbf{RQ1}] To what extent do group and individual fairness evaluation measures differ in their conclusions?
     \item[\textbf{RQ2}] For the same family of measures, how different are the group and individual fairness scores? 
     \item[\textbf{RQ3}] How do different ways of grouping users affect between- and within-group fairness? 
     \item[\textbf{RQ4}] How do between- and within-group fairness relate to individual fairness?
\end{enumerate}

Our results show that group fairness measures often hide unfairness within groups and between individuals, 
highlighting the importance of evaluating fairness beyond the between-group level.

\section{Methodology}
\label{s:setup}

We compare evaluation measures of user-side individual and group fairness in RSs, considering multiple ways of grouping users.

\subsubsection*{Datasets} 
To enable group fairness analysis, three datasets with $\geq 3$ user profile features are selected (see \Cref{tab:stat_and_group} for statistics).

\noindent    \textbf{ML-1M} \cite{Harper2015TheContext}  has 1,000,029 movie ratings (1--5) from 6K users. Users with no/unspecified self-reported gender, age, or occupation are removed, and we exclude users under 18 years to avoid processing the data of minors. We focus on recommending preferred movies, so ratings <3 are discarded, and the levels 4 and 5 are mapped to 1.

\noindent     \textbf{JobRec} \cite{Hamner2012JobChallenge} has 1.6M job applications from 321K users. Given a user's   application history, we focus on recommending job titles that may suit them, keeping only users with information for degree, major, and years of experience. Users with more than 60 years of experience are removed (as this likely indicates erroneous entries). 
    
\noindent      \textbf{LFM-1B} \cite{Schedl2016TheRecommendation} has $1{,}088{,}161{,}692$ music playcounts, from $\sim$120K users. We focus on recommending new track artists for a user to listen to, other than the ones they have listened to in the past, using the dataset after deduplication based on the artist, with 65M interactions (provided by \texttt{RecBole} \cite{Zhao2021RecBole:Algorithms}). The deduplication summarises the total playcount per artist and keeps the last event timestamp. Users without countries, age, or gender information are removed, as are minors (as in ML-1M), and users with age >100 years (as this likely indicates erroneous entries). 

Items without name/title are removed from all datasets. To reduce data sparsity, which may affect LLMRec performance \cite{jiang2025beyond}, we keep users and items with $\geq$ 5 interactions (5-core filtering) for ML-1M and JobRec. We apply 50-core filtering \cite{Makhneva2023MakeSensitive,Wen2023EfficientFiltering,Zhao2023KuaiSim:Systems} to LFM-1B, as it is highly sparse with 5-core filtering. 
The data is temporally split for train/val/test with a ratio of 3:1:1 using a global timeline 
\cite{Meng2020ExploringModels}. From all splits, users and items with $\leq t$ interactions in the train set are removed. A high $t$ can result in very few unique users in the test set, so we choose $t$ such that at least 500 test users remain. For ML-1M and LFM-1B, we set $t=5$ \cite{Xu2024AModels}. For JobRec, we use $t=2$. We remove users/items in the val and test sets that are not in the train set. 

\begin{table}
\centering
\caption{Preprocessed dataset statistics. 
$n_{G_a}$ is the number of groups for sensitive attribute $a$. We exclude empty groups.}

\label{tab:stat_and_group}
    \resizebox{0.95\linewidth}{!}{
    \begin{tabular}{lrrr}
    \toprule
            & ML-1M \cite{Harper2015TheContext} & JobRec \cite{Hamner2012JobChallenge} & LFM-1B \cite{Schedl2016TheRecommendation} \\
    \midrule
    \#interaction (all splits)    & 467,218  & 210,921    &   15,024,267 \\
    \#item (all splits)  &  3,030   & 19,912    &  51,204  \\
    \#user (test set)& 620 & 523  & 16,611     \\
    \midrule
    sensitive attr. \#1 ($n_{G_1}$)      & gender (2)                        &   degree (3)     & gender (2)                        
    \\
    sensitive attr. \#2 ($n_{G_2}$)     & age (3) &   years of experience (3)      & age (3) 
    \\
    sensitive attr. \#3 ($n_{G_3}$)     & occupation (2) & major (6) & country (5) 
    \\
    \midrule
    \#intersectional groups & 12    &  36   &       29     \\
    min--max subgroup size      & 2--279     &  1--70    &       1--4,260       \\
    median subgroup size     & 31    &  7   &       59      \\
    \bottomrule
    \end{tabular}}
\end{table}

\subsubsection*{User grouping} To study group fairness, we cluster users based on their sensitive attributes (see \Cref{tab:stat_and_group} and App.~A.1 in the code repository). Users cannot belong to two groups at the same time, e.g., age<50 and age$\geq$50. 

For ML-1M, we use gender, age, and occupation as sensitive attributes. Gender is used as is. Age is grouped into: 18--24 years, 25--49 years, and $\geq$50 years \cite{OfficeforNationalStatistics2023Age2021}. 
User occupation is grouped into: non-working (student \cite{U.S.BureauofLaborStatistics2024CollegeSummary,Eurostat2018YoungStatistics}, homemaker, retired, and unemployed) and working (14 occupations ranging from farmer to executive \cite{U.S.BureauofLaborStatistics2023MayEstimates}).
    
For JobRec,  we consider the user's academic degree, years of working experience, and study major as the sensitive attributes. Degree is grouped into: high school, college (associate or vocational degree), and university (bachelor's, master's, and PhD). Years of experience are grouped into: $\leq$5 years, >5--10 years, and >10 years. We group study majors into six fields of study, as per \citeauthor{Xu2024AModels}~\cite{Xu2024AModels}, using manual annotation and fuzzy string matching.\footnote{Details are provided in App.~A.1 in the code repository.}
    
For LFM-1B, the sensitive attributes are gender, age, and country. Gender and age are processed as for ML-1M, and the user's country is mapped to the continent.\footnote{We use the country-continent mapping from \url{https://gist.github.com/achuhunkin/6cb1cbceb23395300aa209aad09e6e5d}, and manually group transcontinental countries.} Users from the North/South Americas are grouped together with Antarctica \cite{Xu2024AModels,Gomez2024AMBAR:Recommenders}.

\subsubsection*{LLM-based Recommenders} Recent work has utilised Large Language Models (LLMs) as recommenders (LLMRecs), with promising results \cite{Hou2024LargeSystems}.\footnote{We also experiment with two collaborative filtering RSs, i.e., UserKNN and NeuMF. See also \Cref{fn:similar_res}.}  
Unlike collaborative filtering models, LLMRecs can easily handle fine-grained user attributes (e.g., users' study major, which can be important for job recommendation) with their world knowledge, although including sensitive attributes in the prompt can impact effectiveness and fairness \cite{Deldjoo2025CFaiRLLM,Xu2024AModels,Zhang2023IsRecommendation}. 
We therefore study the effectiveness and fairness of LLMRecs under few-shot learning. 
To ensure comparable performance, we use four open-source, similar-sized LLMs released in July--Nov'24: 
Llama-3.1-8B-Instruct \cite{grattafiori2024llama3herdmodels}, Qwen2.5-7B-Instruct \cite{qwen2.5}, GLM-4-9B-chat \cite{glm2024chatglm}, and Ministral-8B-Instruct-2410 \cite{Mistral}. 
The temperature is fixed at 0 for each LLM to obtain deterministic output.

The LLMs are prompted using in-context learning (ICL) 
strategy \cite{Hou2024LargeSystems}, as this has been shown to outperform sequential and recency-focused prompting.\footnote{The full prompt templates and examples are provided in App.~A.2.}
We only prompt for users that exist in the test set, as otherwise it is not possible to evaluate the recommendation effectiveness. 
In the prompt, we provide the users' train items as the interaction history and the val items as the few-shot samples. To guide the recommendation generation with the ICL strategy, val items are used as examples of what should be recommended to a user, considering their historical interactions. 
A maximum of 10 most recent train and val items each are provided, as having too many items in the prompt may reduce effectiveness \cite{jiang2025beyond, Hou2024LargeSystems}. 
To avoid inflated performance, we do not prompt with a sampled candidate item list \cite{Krichene2022OnRecommendation}. Instead, we add restriction in the prompt to narrow down the item search space, e.g., for ML-1M, the movies should be between certain years, based on the movie release year in the metadata file. For LFM-1B, the prompt also includes the playcount, which is important in music recommendation \cite{Gomez2024AMBAR:Recommenders}.

Based on the inclusion/exclusion of user sensitive attributes, we create two prompt types \cite{Deldjoo2025CFaiRLLM, Zhang2023IsRecommendation, Tommasel2024FairnessRecommendations,deldjoo2024normativeframeworkbenchmarkingconsumer}: Sensitive (S), which has both interaction history and all three sensitive attributes, and Non-Sensitive (NS), which has only the interaction history.

To evaluate LLMRecs, we perform fuzzy string matching between the list of recommended items and item names in the test set \cite{Deldjoo2024UnderstandingRecency,jiang2025beyond,He2023LLMConversational,liang-etal-2024-llm,dipalma2024evaluatingchatgptrecommendersystem}, by using the TF-IDF \cite{Jones1972ARetrieval} of the items' character-based n-gram \cite{Lian2024RecAI:Systems}.\footnote{Metrics based on n-gram have been used to evaluate the performance of (conversational) RSs \cite{ravaut-etal-2024-parameter}.}
If the item name similarity exceeds a pre-set threshold ($\geq0.75$), we count it as a match (i.e., the LLMRec successfully recommends an item that exists in a user's test set).\footnote{To our knowledge, no existing work has evaluated the effect of various similarity thresholds for this context.} Our LLMRecs experiments are carried out with \texttt{vllm} \cite{Kwon2023EfficientPagedAttention} and \texttt{RecLM-eval} \cite{Lian2024RecAI:Systems}. 

\subsubsection*{Evaluation} 

Recommendation effectiveness (\eff{}) and fairness (\fair{}) are measured at $k=10$ for all LLMRecs. For \eff{}, the mean Hit Rate (HR), MRR, P@$k$ (P), and NDCG@$k$ \cite{Jarvelin2002NDCG} are computed over all users. Group and individual \fair{} measures are computed in two steps: first, computing an \eff{} score per user/group as a `base score'; and second, aggregating the `base score' between users/groups with a \fair{} measure. P and NDCG are used as base scores to represent set- and rank-based measures. 

\noindent\textbf{Group fairness.} 
We compute all existing fairness measures for two or more user groups in RSs,\footnote{Measures that can only be used for exactly two groups are excluded.} 
that are published up to March 2025: Average scores of the worst 25\% groups (Min \cite{Wang2024IntersectionalRecommendation}), Range \cite{Liu2024SelfAdapative}, SD \cite{Zhang2023IsRecommendation,Liu2024SelfAdapative},  
MAD \cite{Fu2020FairnessGraph}, Gini \cite{Pastor2024IntersectionalDivergence,Ferraro2024GenderImbalance,ghosh2024reducingpopulationlevelinequalityimprove}, CV \cite{Zhu2020MeasuringSystems}, 
FStat \cite{Wan2020AddressingRecommendations}, KL \cite{Amigo2023ASystems}, and GCE \cite{Deldjoo2021ASystems,Deldjoo2019RecommenderEntropy}. 
We also compute the Atkinson Index (Atk \cite{Atkinson1970OnInequality}), an income inequality measure that considers within-group variations.\footnote{This measure can be transformed into Generalised Entropy \cite{Speicher2018UnifiedIndices,Shorrocks1980ClassDecomposable}.} 
The between- and within-group fairness version of the measures are denoted as $\cdot_\text{b-group}$ and $\cdot_\text{w-group}$ respectively.\footnote{The term \textit{group fairness} refers to between-group fairness; the latter is used when we compare fairness between and within groups.} 
We provide the measure formulations and technical details in App.~A.3.

\noindent\textbf{Individual fairness.} 
Fairness for individual users is quantified with SD \cite{Patro2020FairRec:Platforms}, Gini \cite{Leonhardt2018UserSystems}, and Atk. The subscript $\cdot$\ind{} indicates the individual fairness version of the measure. 
SD and Gini are the only \fair{} measures that have been used for both individual and group user fairness, while Atk\ind{} can be decomposed into between- and within-group fairness with no residuals \cite{Blackorby1999,Dayioğlu02006ImputedIndex,Bourguignon1979DecomposableMeasures}. 
While other group \fair{} measures can also be used to measure individual fairness, their scores may not be informative, e.g., Min may be zero for most models, as having most users scoring P=0 or NDCG=0 is common. 

\section{Empirical Analysis}
\label{s:experiment}

\subsubsection*{Evaluation of all LLMRecs}

\begin{table}
    \caption{
 Effectiveness (\eff{}) and fairness (\fair{}) scores at cut-off $k=10$ for intersectional groups (\textsc{Grp.}) and for individuals (\textsc{Ind.}) of LLMRecs with non-sensitive (NS) and sensitive (S) prompts. All \fair{} scores are computed with NDCG. All measures range in [0,1], except for the \textsc{Grp.} measures below the grey lines. The best \textsc{Eff}/\fair{} scores are bolded. 
 Darker green marks scores closer to the best \eff{}/\fair{} per measure. 
 $\uparrow/\downarrow$ means the higher/lower the better. 
 }
    \label{tab:col_LLM-NDCG}
    \resizebox{0.98\columnwidth}{!}{
\begin{tabular}{ll*{2}{r}|*{2}{r}|*{2}{r}|*{2}{r}}
\toprule
\toprule
 &  LLMRec& \multicolumn{2}{c|}{GLM-4-9B} & \multicolumn{2}{c|}{Llama-3.1-8B} & \multicolumn{2}{c|}{Ministral-8B} & \multicolumn{2}{c}{Qwen2.5-7B} \\ 
\midrule
 &  prompt type & NS & S & NS & S & NS & S & NS & S \\
\midrule
\midrule
 &  & \multicolumn{8}{c}{ML-1M} \\ 
 \midrule
\multirow[c]{3}{*}{\rotatebox[origin=c]{90}{\textsc{Eff}}} & $\uparrow$ HR & \bfseries {\cellcolor[HTML]{54A075}}  0.377 & {\cellcolor[HTML]{6CAD88}}  0.358 & {\cellcolor[HTML]{ECF2EE}}  0.260 & {\cellcolor[HTML]{E1ECE5}}  0.269 & {\cellcolor[HTML]{81B899}}  0.342 & {\cellcolor[HTML]{84BA9B}}  0.340 & {\cellcolor[HTML]{66AA83}}  0.363 & {\cellcolor[HTML]{5CA47B}}  0.371 \\
 & $\uparrow$ MRR & \bfseries {\cellcolor[HTML]{54A075}}  0.189 & {\cellcolor[HTML]{6EAE89}}  0.174 & {\cellcolor[HTML]{ECF2EE}}  0.101 & {\cellcolor[HTML]{D8E7DE}}  0.113 & {\cellcolor[HTML]{88BC9E}}  0.159 & {\cellcolor[HTML]{6FAF8B}}  0.173 & {\cellcolor[HTML]{7DB696}}  0.165 & {\cellcolor[HTML]{64A881}}  0.180 \\
 & $\uparrow$ NDCG & \bfseries {\cellcolor[HTML]{54A075}}  0.231 & {\cellcolor[HTML]{6FAE8A}}  0.215 & {\cellcolor[HTML]{ECF2EE}}  0.140 & {\cellcolor[HTML]{DFEBE4}}  0.148 & {\cellcolor[HTML]{88BC9E}}  0.200 & {\cellcolor[HTML]{7AB593}}  0.208 & {\cellcolor[HTML]{7CB695}}  0.207 & {\cellcolor[HTML]{65A982}}  0.221 \\
\cline{1-10}
\multirow[c]{10}{*}{\rotatebox[origin=c]{90}{\textsc{Fair (Grp.)}}} & $\uparrow$ Min & \bfseries {\cellcolor[HTML]{54A075}}  0.166 & {\cellcolor[HTML]{7DB696}}  0.137 & {\cellcolor[HTML]{C6DED0}}  0.086 & {\cellcolor[HTML]{A6CCB7}}  0.108 & {\cellcolor[HTML]{ECF2EE}}  0.059 & {\cellcolor[HTML]{A8CDB8}}  0.107 & {\cellcolor[HTML]{E0EBE4}}  0.068 & {\cellcolor[HTML]{BFDACA}}  0.091 \\
 & $\downarrow$ Range & \bfseries {\cellcolor[HTML]{54A075}}  0.188 & {\cellcolor[HTML]{5EA57D}}  0.208 & {\cellcolor[HTML]{A6CCB6}}  0.356 & {\cellcolor[HTML]{B0D2BE}}  0.376 & {\cellcolor[HTML]{86BB9C}}  0.290 & {\cellcolor[HTML]{ECF2EE}}  0.500 & {\cellcolor[HTML]{98C5AB}}  0.328 & {\cellcolor[HTML]{7EB796}}  0.274 \\
 & $\downarrow$ SD & \bfseries {\cellcolor[HTML]{54A075}}  0.055 & {\cellcolor[HTML]{61A77F}}  0.061 & {\cellcolor[HTML]{9CC7AE}}  0.088 & {\cellcolor[HTML]{9CC7AE}}  0.088 & {\cellcolor[HTML]{95C3A9}}  0.085 & {\cellcolor[HTML]{ECF2EE}}  0.125 & {\cellcolor[HTML]{A6CCB7}}  0.093 & {\cellcolor[HTML]{84BA9B}}  0.077 \\
 & $\downarrow$ MAD & \bfseries {\cellcolor[HTML]{54A075}}  0.067 & {\cellcolor[HTML]{66AA83}}  0.076 & {\cellcolor[HTML]{84BA9B}}  0.091 & {\cellcolor[HTML]{7DB695}}  0.087 & {\cellcolor[HTML]{95C3A9}}  0.099 & {\cellcolor[HTML]{ECF2EE}}  0.142 & {\cellcolor[HTML]{AFD1BE}}  0.112 & {\cellcolor[HTML]{83B99A}}  0.090 \\
 & $\downarrow$ Gini & \bfseries {\cellcolor[HTML]{54A075}}  0.130 & {\cellcolor[HTML]{74B18F}}  0.161 & {\cellcolor[HTML]{D7E7DE}}  0.255 & {\cellcolor[HTML]{B9D6C5}}  0.226 & {\cellcolor[HTML]{D0E3D8}}  0.248 & {\cellcolor[HTML]{ECF2EE}}  0.275 & {\cellcolor[HTML]{D7E7DE}}  0.255 & {\cellcolor[HTML]{98C5AB}}  0.195 \\
 & $\downarrow$ Atk & \bfseries {\cellcolor[HTML]{54A075}}  0.015 & {\cellcolor[HTML]{68AB85}}  0.020 & {\cellcolor[HTML]{A8CDB8}}  0.036 & {\cellcolor[HTML]{7CB695}}  0.025 & {\cellcolor[HTML]{9CC7AE}}  0.033 & {\cellcolor[HTML]{DDEAE2}}  0.049 & {\cellcolor[HTML]{ECF2EE}}  0.053 & {\cellcolor[HTML]{B4D4C2}}  0.039 \\
\arrayrulecolor{gray!50}\cline{2-10}\arrayrulecolor{black}
 & $\downarrow$ CV & \bfseries {\cellcolor[HTML]{54A075}}  0.233 & {\cellcolor[HTML]{6EAE89}}  0.285 & {\cellcolor[HTML]{ECF2EE}}  0.540 & {\cellcolor[HTML]{DAE8DF}}  0.502 & {\cellcolor[HTML]{C7DED1}}  0.465 & {\cellcolor[HTML]{E5EEE8}}  0.525 & {\cellcolor[HTML]{C5DDCF}}  0.460 & {\cellcolor[HTML]{96C3A9}}  0.365 \\
 & $\downarrow$ FStat & \bfseries {\cellcolor[HTML]{54A075}}  0.468 & {\cellcolor[HTML]{74B18E}}  0.714 & {\cellcolor[HTML]{84BA9B}}  0.841 & {\cellcolor[HTML]{6CAD88}}  0.654 & {\cellcolor[HTML]{7BB594}}  0.767 & {\cellcolor[HTML]{BDD9C9}}  1.278 & {\cellcolor[HTML]{ECF2EE}}  1.645 & {\cellcolor[HTML]{B5D5C2}}  1.220 \\
 & $\downarrow$ KL & {\cellcolor[HTML]{84BA9B}}  1.121 & {\cellcolor[HTML]{87BC9E}}  1.138 & {\cellcolor[HTML]{ECF2EE}}  1.674 & {\cellcolor[HTML]{E6EFE9}}  1.640 & \bfseries {\cellcolor[HTML]{54A075}}  0.866 & {\cellcolor[HTML]{ECF2EE}}  1.671 & {\cellcolor[HTML]{93C2A7}}  1.198 & {\cellcolor[HTML]{79B492}}  1.063 \\
 & $\downarrow$ GCE & \bfseries {\cellcolor[HTML]{54A075}}  0.028 & {\cellcolor[HTML]{54A075}}  0.050 & {\cellcolor[HTML]{54A075}}  0.112 & {\cellcolor[HTML]{54A075}}  0.104 & {\cellcolor[HTML]{ECF2EE}}  659.844 & {\cellcolor[HTML]{ECF2EE}}  659.741 & {\cellcolor[HTML]{54A075}}  0.239 & {\cellcolor[HTML]{54A075}}  0.198 \\
\cline{1-10}
\multirow[c]{3}{*}{\rotatebox[origin=c]{90}{\textsc{Fair}} \rotatebox[origin=c]{90}{\textsc{(Ind.)}}}  & $\downarrow$ SD & {\cellcolor[HTML]{ECF2EE}}  0.330 & {\cellcolor[HTML]{D6E6DD}}  0.320 & \bfseries {\cellcolor[HTML]{54A075}}  0.262 & {\cellcolor[HTML]{6AAC87}}  0.272 & {\cellcolor[HTML]{BDD9C9}}  0.309 & {\cellcolor[HTML]{DAE8E0}}  0.322 & {\cellcolor[HTML]{BBD8C7}}  0.308 & {\cellcolor[HTML]{DFEBE4}}  0.324 \\
 & $\downarrow$ Gini & \bfseries {\cellcolor[HTML]{54A075}}  0.705 & {\cellcolor[HTML]{6EAE89}}  0.721 & {\cellcolor[HTML]{ECF2EE}}  0.799 & {\cellcolor[HTML]{E3EDE7}}  0.793 & {\cellcolor[HTML]{86BB9D}}  0.736 & {\cellcolor[HTML]{86BB9D}}  0.736 & {\cellcolor[HTML]{6EAE89}}  0.721 & {\cellcolor[HTML]{64A982}}  0.715 \\
 & $\downarrow$ Atk & \bfseries {\cellcolor[HTML]{54A075}}  0.636 & {\cellcolor[HTML]{6DAE89}}  0.655 & {\cellcolor[HTML]{ECF2EE}}  0.751 & {\cellcolor[HTML]{E0ECE5}}  0.742 & {\cellcolor[HTML]{84BA9B}}  0.672 & {\cellcolor[HTML]{85BA9C}}  0.673 & {\cellcolor[HTML]{69AB86}}  0.652 & {\cellcolor[HTML]{5EA57D}}  0.644 \\
\midrule
\midrule
 &  & \multicolumn{8}{c}{JobRec} \\ 
\midrule
\multirow[c]{3}{*}{\rotatebox[origin=c]{90}{\textsc{Eff}}} & $\uparrow$ HR & {\cellcolor[HTML]{7AB493}}  0.054 & {\cellcolor[HTML]{C3DCCD}}  0.033 & {\cellcolor[HTML]{9DC7AF}}  0.044 & {\cellcolor[HTML]{ECF2EE}}  0.021 & {\cellcolor[HTML]{6FAF8B}}  0.057 & {\cellcolor[HTML]{9DC7AF}}  0.044 & \bfseries {\cellcolor[HTML]{54A075}}  0.065 & {\cellcolor[HTML]{6FAF8B}}  0.057 \\
 & $\uparrow$ MRR & {\cellcolor[HTML]{7FB797}}  0.037 & {\cellcolor[HTML]{B6D5C3}}  0.023 & {\cellcolor[HTML]{CDE1D5}}  0.017 & {\cellcolor[HTML]{ECF2EE}}  0.009 & \bfseries {\cellcolor[HTML]{54A075}}  0.048 & {\cellcolor[HTML]{8EC0A4}}  0.033 & {\cellcolor[HTML]{5CA47B}}  0.046 & {\cellcolor[HTML]{5FA67E}}  0.045 \\
 & $\uparrow$ NDCG & {\cellcolor[HTML]{78B392}}  0.041 & {\cellcolor[HTML]{B8D6C5}}  0.025 & {\cellcolor[HTML]{BCD8C8}}  0.024 & {\cellcolor[HTML]{ECF2EE}}  0.012 & \bfseries {\cellcolor[HTML]{54A075}}  0.050 & {\cellcolor[HTML]{8CBEA2}}  0.036 & \bfseries {\cellcolor[HTML]{54A075}}  0.050 & {\cellcolor[HTML]{5CA47B}}  0.048 \\
\cline{1-10}
\multirow[c]{10}{*}{\rotatebox[origin=c]{90}{\textsc{Fair (Grp.)}}} & $\uparrow$ Min & \bfseries {\cellcolor[HTML]{ECF2EE}}  0.000 & \bfseries {\cellcolor[HTML]{ECF2EE}}  0.000 & \bfseries {\cellcolor[HTML]{ECF2EE}}  0.000 & \bfseries {\cellcolor[HTML]{ECF2EE}}  0.000 & \bfseries {\cellcolor[HTML]{ECF2EE}}  0.000 & \bfseries {\cellcolor[HTML]{ECF2EE}}  0.000 & \bfseries {\cellcolor[HTML]{ECF2EE}}  0.000 & \bfseries {\cellcolor[HTML]{ECF2EE}}  0.000 \\
 & $\downarrow$ Range & {\cellcolor[HTML]{ECF2EE}}  0.500 & \bfseries {\cellcolor[HTML]{54A075}}  0.083 & {\cellcolor[HTML]{ECF2EE}}  0.500 & {\cellcolor[HTML]{ECF2EE}}  0.500 & {\cellcolor[HTML]{AFD1BE}}  0.333 & {\cellcolor[HTML]{74B18E}}  0.170 & {\cellcolor[HTML]{AFD1BE}}  0.333 & {\cellcolor[HTML]{AFD1BE}}  0.333 \\
 & $\downarrow$ SD & {\cellcolor[HTML]{ECF2EE}}  0.093 & \bfseries {\cellcolor[HTML]{54A075}}  0.023 & {\cellcolor[HTML]{DDEAE2}}  0.086 & {\cellcolor[HTML]{D4E5DB}}  0.082 & {\cellcolor[HTML]{C3DCCD}}  0.074 & {\cellcolor[HTML]{7DB696}}  0.042 & {\cellcolor[HTML]{BFDACA}}  0.072 & {\cellcolor[HTML]{BCD8C8}}  0.071 \\
 & $\downarrow$ MAD & {\cellcolor[HTML]{ECF2EE}}  0.066 & \bfseries {\cellcolor[HTML]{54A075}}  0.019 & {\cellcolor[HTML]{C2DCCD}}  0.053 & {\cellcolor[HTML]{88BC9E}}  0.035 & {\cellcolor[HTML]{E9F1EC}}  0.065 & {\cellcolor[HTML]{88BC9E}}  0.035 & {\cellcolor[HTML]{E6EFE9}}  0.064 & {\cellcolor[HTML]{CCE1D5}}  0.056 \\
 & $\downarrow$ Gini & {\cellcolor[HTML]{97C4AB}}  0.828 & {\cellcolor[HTML]{9CC7AE}}  0.834 & {\cellcolor[HTML]{ADD0BC}}  0.857 & {\cellcolor[HTML]{ECF2EE}}  0.939 & {\cellcolor[HTML]{61A77F}}  0.757 & {\cellcolor[HTML]{7AB593}}  0.790 & \bfseries {\cellcolor[HTML]{54A075}}  0.740 & {\cellcolor[HTML]{89BC9F}}  0.809 \\
 & $\downarrow$ Atk & {\cellcolor[HTML]{73B18E}}  0.547 & {\cellcolor[HTML]{66AA84}}  0.518 & {\cellcolor[HTML]{AACEB9}}  0.677 & {\cellcolor[HTML]{ECF2EE}}  0.834 & {\cellcolor[HTML]{68AB85}}  0.522 & {\cellcolor[HTML]{5FA67E}}  0.499 & \bfseries {\cellcolor[HTML]{54A075}}  0.473 & {\cellcolor[HTML]{5FA67E}}  0.500 \\
\arrayrulecolor{gray!50}\cline{2-10}\arrayrulecolor{black}
 & $\downarrow$ CV & {\cellcolor[HTML]{78B492}}  2.396 & {\cellcolor[HTML]{69AB86}}  2.102 & {\cellcolor[HTML]{93C2A7}}  2.886 & {\cellcolor[HTML]{ECF2EE}}  4.549 & {\cellcolor[HTML]{56A177}}  1.760 & {\cellcolor[HTML]{5FA67E}}  1.928 & \bfseries {\cellcolor[HTML]{54A075}}  1.709 & {\cellcolor[HTML]{69AB86}}  2.107 \\
 & $\downarrow$ FStat & {\cellcolor[HTML]{83B99A}}  1.035 & \bfseries {\cellcolor[HTML]{54A075}}  0.546 & {\cellcolor[HTML]{BAD7C6}}  1.613 & {\cellcolor[HTML]{ECF2EE}}  2.137 & {\cellcolor[HTML]{80B898}}  1.005 & {\cellcolor[HTML]{5AA37A}}  0.611 & {\cellcolor[HTML]{77B391}}  0.908 & {\cellcolor[HTML]{78B492}}  0.928 \\
 & $\downarrow$ KL & {\cellcolor[HTML]{91C1A6}}  3.218 & \bfseries {\cellcolor[HTML]{54A075}}  1.428 & {\cellcolor[HTML]{ACCFBB}}  3.979 & {\cellcolor[HTML]{ECF2EE}}  5.861 & {\cellcolor[HTML]{64A881}}  1.881 & {\cellcolor[HTML]{5FA67E}}  1.754 & {\cellcolor[HTML]{6AAC87}}  2.069 & {\cellcolor[HTML]{71B08C}}  2.288 \\
 & $\downarrow$ GCE & {\cellcolor[HTML]{80B898}}  1685.926 & {\cellcolor[HTML]{D7E7DD}}  1979.103 & {\cellcolor[HTML]{80B898}}  1685.994 & {\cellcolor[HTML]{ECF2EE}}  2052.498 & {\cellcolor[HTML]{69AC86}}  1612.574 & {\cellcolor[HTML]{95C3A9}}  1759.177 & \bfseries {\cellcolor[HTML]{54A075}}  1539.278 & {\cellcolor[HTML]{95C3A9}}  1759.185 \\
\cline{1-10}
\multirow[c]{3}{*}{\rotatebox[origin=c]{90}{\textsc{Fair}} \rotatebox[origin=c]{90}{\textsc{(Ind.)}}} & $\downarrow$ SD & {\cellcolor[HTML]{C9DFD2}}  0.183 & {\cellcolor[HTML]{9CC7AE}}  0.147 & {\cellcolor[HTML]{7BB594}}  0.121 & \bfseries {\cellcolor[HTML]{54A075}}  0.090 & {\cellcolor[HTML]{ECF2EE}}  0.211 & {\cellcolor[HTML]{BED9C9}}  0.174 & {\cellcolor[HTML]{E4EEE8}}  0.204 & {\cellcolor[HTML]{E3EDE7}}  0.203 \\
 & $\downarrow$ Gini & {\cellcolor[HTML]{78B392}}  0.956 & {\cellcolor[HTML]{C0DACB}}  0.974 & {\cellcolor[HTML]{A0C9B2}}  0.966 & {\cellcolor[HTML]{ECF2EE}}  0.985 & {\cellcolor[HTML]{5CA47B}}  0.949 & {\cellcolor[HTML]{94C2A8}}  0.963 & \bfseries {\cellcolor[HTML]{54A075}}  0.947 & {\cellcolor[HTML]{64A881}}  0.951 \\
 & $\downarrow$ Atk & {\cellcolor[HTML]{7BB594}}  0.948 & {\cellcolor[HTML]{C5DDCF}}  0.969 & {\cellcolor[HTML]{9FC8B0}}  0.958 & {\cellcolor[HTML]{ECF2EE}}  0.980 & {\cellcolor[HTML]{6CAD88}}  0.944 & {\cellcolor[HTML]{9BC6AE}}  0.957 & \bfseries {\cellcolor[HTML]{54A075}}  0.937 & {\cellcolor[HTML]{6CAD88}}  0.944 \\
\midrule
\midrule
 &  & \multicolumn{8}{c}{LFM-1B} \\ 
\midrule
\multirow[c]{3}{*}{\rotatebox[origin=c]{90}{\textsc{Eff}}} & $\uparrow$ HR & {\cellcolor[HTML]{55A176}}  0.658 & \bfseries {\cellcolor[HTML]{54A075}}  0.661 & {\cellcolor[HTML]{6BAC87}}  0.609 & {\cellcolor[HTML]{67AA84}}  0.618 & {\cellcolor[HTML]{B1D2BF}}  0.451 & {\cellcolor[HTML]{ECF2EE}}  0.317 & {\cellcolor[HTML]{D2E4DA}}  0.375 & {\cellcolor[HTML]{ECF2EE}}  0.317 \\
 & $\uparrow$ MRR & \bfseries {\cellcolor[HTML]{54A075}}  0.409 & {\cellcolor[HTML]{55A075}}  0.408 & {\cellcolor[HTML]{7AB493}}  0.347 & {\cellcolor[HTML]{74B18E}}  0.357 & {\cellcolor[HTML]{ACCFBB}}  0.266 & {\cellcolor[HTML]{E4EEE8}}  0.174 & {\cellcolor[HTML]{D4E5DB}}  0.199 & {\cellcolor[HTML]{ECF2EE}}  0.160 \\
 & $\uparrow$ NDCG & {\cellcolor[HTML]{54A075}}  0.462 & \bfseries {\cellcolor[HTML]{54A075}}  0.463 & {\cellcolor[HTML]{75B28F}}  0.406 & {\cellcolor[HTML]{6FAF8B}}  0.415 & {\cellcolor[HTML]{ACCFBB}}  0.310 & {\cellcolor[HTML]{E7EFEA}}  0.208 & {\cellcolor[HTML]{D4E5DB}}  0.241 & {\cellcolor[HTML]{ECF2EE}}  0.198 \\
\cline{1-10}
\multirow[c]{10}{*}{\rotatebox[origin=c]{90}{\textsc{Fair (Grp.)}}} & $\uparrow$ Min & {\cellcolor[HTML]{69AC86}}  0.240 & \bfseries {\cellcolor[HTML]{54A075}}  0.273 & {\cellcolor[HTML]{5AA37A}}  0.264 & {\cellcolor[HTML]{59A379}}  0.265 & {\cellcolor[HTML]{C2DBCC}}  0.107 & {\cellcolor[HTML]{E3EDE7}}  0.058 & {\cellcolor[HTML]{CDE1D5}}  0.090 & {\cellcolor[HTML]{ECF2EE}}  0.043 \\
 & $\downarrow$ Range & {\cellcolor[HTML]{92C1A6}}  0.604 & {\cellcolor[HTML]{92C1A6}}  0.604 & {\cellcolor[HTML]{D2E4D9}}  0.884 & {\cellcolor[HTML]{DDEAE2}}  0.931 & {\cellcolor[HTML]{EBF1ED}}  0.993 & {\cellcolor[HTML]{ECF2EE}}  1.000 & {\cellcolor[HTML]{80B898}}  0.525 & \bfseries {\cellcolor[HTML]{54A075}}  0.331 \\
 & $\downarrow$ SD & {\cellcolor[HTML]{B8D6C5}}  0.149 & {\cellcolor[HTML]{97C4AB}}  0.130 & {\cellcolor[HTML]{C8DED1}}  0.158 & {\cellcolor[HTML]{A9CEB8}}  0.140 & {\cellcolor[HTML]{ECF2EE}}  0.179 & {\cellcolor[HTML]{EBF1ED}}  0.178 & {\cellcolor[HTML]{81B899}}  0.117 & \bfseries {\cellcolor[HTML]{54A075}}  0.091 \\
 & $\downarrow$ MAD & {\cellcolor[HTML]{A5CBB5}}  0.138 & {\cellcolor[HTML]{7DB696}}  0.121 & {\cellcolor[HTML]{D3E5DA}}  0.158 & {\cellcolor[HTML]{92C1A6}}  0.130 & {\cellcolor[HTML]{ECF2EE}}  0.169 & {\cellcolor[HTML]{D1E3D8}}  0.157 & {\cellcolor[HTML]{87BB9D}}  0.125 & \bfseries {\cellcolor[HTML]{54A075}}  0.103 \\
 & $\downarrow$ Gini & {\cellcolor[HTML]{64A881}}  0.162 & \bfseries {\cellcolor[HTML]{54A075}}  0.139 & {\cellcolor[HTML]{73B18E}}  0.184 & {\cellcolor[HTML]{63A881}}  0.161 & {\cellcolor[HTML]{B9D7C6}}  0.285 & {\cellcolor[HTML]{ECF2EE}}  0.358 & {\cellcolor[HTML]{A9CEB9}}  0.262 & {\cellcolor[HTML]{C2DCCD}}  0.298 \\
 & $\downarrow$ Atk & \bfseries {\cellcolor[HTML]{54A075}}  0.002 & \bfseries {\cellcolor[HTML]{54A075}}  0.002 & {\cellcolor[HTML]{5EA57D}}  0.003 & {\cellcolor[HTML]{68AB85}}  0.004 & {\cellcolor[HTML]{7DB695}}  0.006 & {\cellcolor[HTML]{C4DCCE}}  0.013 & {\cellcolor[HTML]{87BB9D}}  0.007 & {\cellcolor[HTML]{ECF2EE}}  0.017 \\
\arrayrulecolor{gray!50}\cline{2-10}\arrayrulecolor{black}
 & $\downarrow$ CV & {\cellcolor[HTML]{63A881}}  0.363 & \bfseries {\cellcolor[HTML]{54A075}}  0.309 & {\cellcolor[HTML]{69AB86}}  0.383 & {\cellcolor[HTML]{61A77F}}  0.356 & {\cellcolor[HTML]{AFD1BD}}  0.625 & {\cellcolor[HTML]{ECF2EE}}  0.841 & {\cellcolor[HTML]{8DBFA3}}  0.510 & {\cellcolor[HTML]{97C4AB}}  0.545 \\
 & $\downarrow$ FStat & {\cellcolor[HTML]{6AAC87}}  2.188 & \bfseries {\cellcolor[HTML]{54A075}}  1.809 & {\cellcolor[HTML]{92C1A6}}  2.859 & {\cellcolor[HTML]{CCE1D5}}  3.841 & {\cellcolor[HTML]{91C1A6}}  2.855 & {\cellcolor[HTML]{B1D2BF}}  3.382 & {\cellcolor[HTML]{73B18E}}  2.337 & {\cellcolor[HTML]{ECF2EE}}  4.385 \\
 & $\downarrow$ KL & {\cellcolor[HTML]{7EB796}}  2.645 & {\cellcolor[HTML]{90C0A5}}  2.783 & {\cellcolor[HTML]{EBF1ED}}  3.497 & {\cellcolor[HTML]{C4DCCE}}  3.189 & {\cellcolor[HTML]{ADD0BC}}  3.011 & {\cellcolor[HTML]{ECF2EE}}  3.508 & {\cellcolor[HTML]{9CC7AE}}  2.878 & \bfseries {\cellcolor[HTML]{54A075}}  2.318 \\
 & $\downarrow$ GCE & {\cellcolor[HTML]{A0C9B1}}  338.842 & {\cellcolor[HTML]{7AB493}}  225.899 & {\cellcolor[HTML]{54A075}}  112.987 & \bfseries {\cellcolor[HTML]{54A075}}  112.974 & {\cellcolor[HTML]{C7DED0}}  451.826 & {\cellcolor[HTML]{C7DED0}}  451.875 & {\cellcolor[HTML]{C7DED0}}  451.822 & {\cellcolor[HTML]{ECF2EE}}  564.765 \\
\cline{1-10}
\multirow[c]{3}{*}{\rotatebox[origin=c]{90}{\textsc{Fair}} \rotatebox[origin=c]{90}{\textsc{(Ind.)}}}  & $\downarrow$ SD & {\cellcolor[HTML]{ECF2EE}}  0.380 & {\cellcolor[HTML]{E7F0EA}}  0.378 & {\cellcolor[HTML]{D3E5DA}}  0.370 & {\cellcolor[HTML]{D5E6DC}}  0.371 & {\cellcolor[HTML]{E5EEE8}}  0.377 & {\cellcolor[HTML]{77B391}}  0.334 & {\cellcolor[HTML]{8BBEA1}}  0.342 & \bfseries {\cellcolor[HTML]{54A075}}  0.320 \\
 & $\downarrow$ Gini & {\cellcolor[HTML]{54A075}}  0.462 & \bfseries {\cellcolor[HTML]{54A075}}  0.461 & {\cellcolor[HTML]{6DAE89}}  0.510 & {\cellcolor[HTML]{69AB86}}  0.502 & {\cellcolor[HTML]{B0D2BF}}  0.638 & {\cellcolor[HTML]{EBF1ED}}  0.750 & {\cellcolor[HTML]{D2E4DA}}  0.703 & {\cellcolor[HTML]{ECF2EE}}  0.753 \\
 & $\downarrow$ Atk & {\cellcolor[HTML]{55A176}}  0.361 & \bfseries {\cellcolor[HTML]{54A075}}  0.358 & {\cellcolor[HTML]{6BAC87}}  0.409 & {\cellcolor[HTML]{66AA84}}  0.400 & {\cellcolor[HTML]{B0D2BF}}  0.563 & {\cellcolor[HTML]{ECF2EE}}  0.694 & {\cellcolor[HTML]{D2E4DA}}  0.638 & {\cellcolor[HTML]{ECF2EE}}  0.695 \\
\bottomrule
\bottomrule
\end{tabular}}
\end{table}

\eff{} scores and NDCG-based \fair{} scores for group fairness and individual fairness of LLMRecs are shown in \Cref{tab:col_LLM-NDCG}; 
 the P-based \fair{} measures are presented in App.~B.1 as they show similar results. 
 We evaluate group fairness considering the intersectionality of all three sensitive user attributes.\footnote{We find similar results for collaborative filtering-based RSs. The results were omitted for brevity and to focus on LLMRecs. \label{fn:similar_res}}
 
\noindent\textbf{\eff{} and \fair{} scores}. GLM-4-9B has the best \eff{} and \fair{} scores for all datasets, except for JobRec, where all LLMRecs have equally low \eff{}. This could be because the job titles are very specific and contain more noise than movie/titles and artist names (e.g., `Chicago Concierge Manager', `Account Executive - 40k - 70K per year'). Generally, as group/individual \fair{} measures, Gini, Atk, and CV are fairer when NDCG is better; this is because their equations include division by the mean NDCG. As such, these measures can distinguish group fairness between two models that have similar SD$_\text{b-group}$/SD$_\text{ind}$ but differ in NDCG (e.g., the sensitive and non-sensitive prompts of Ministral-8B for LFM-1B), or vice versa. 

\noindent\textbf{Comparison of sensitive and non-sensitive prompts}. The \eff{} and \fair{} scores between sensitive (S) and non-sensitive (NS) prompts are generally comparable, implying that including sensitive attributes in the prompt has little effect on effectiveness/fairness. An exception to this is Ministral-8B for LFM-1B, which may be due to the LLM's relatively high gender bias \cite{cohere2025commandaenterprisereadylarge}.

\noindent\textbf{Group vs.~individual fairness}. For Atk/Gini, the fairest model for intersectional groups and for individual users are always the same, but this is not the case for SD. Thus, if sensitive attribute information is missing, Atk$_\text{ind}$/Gini$_\text{ind}$ can estimate which model is the fairest for intersectional groups.

\subsubsection*{Agreement of group and individual fairness measures (RQ1)}

\begin{figure}
    \includegraphics[width=0.95\linewidth]{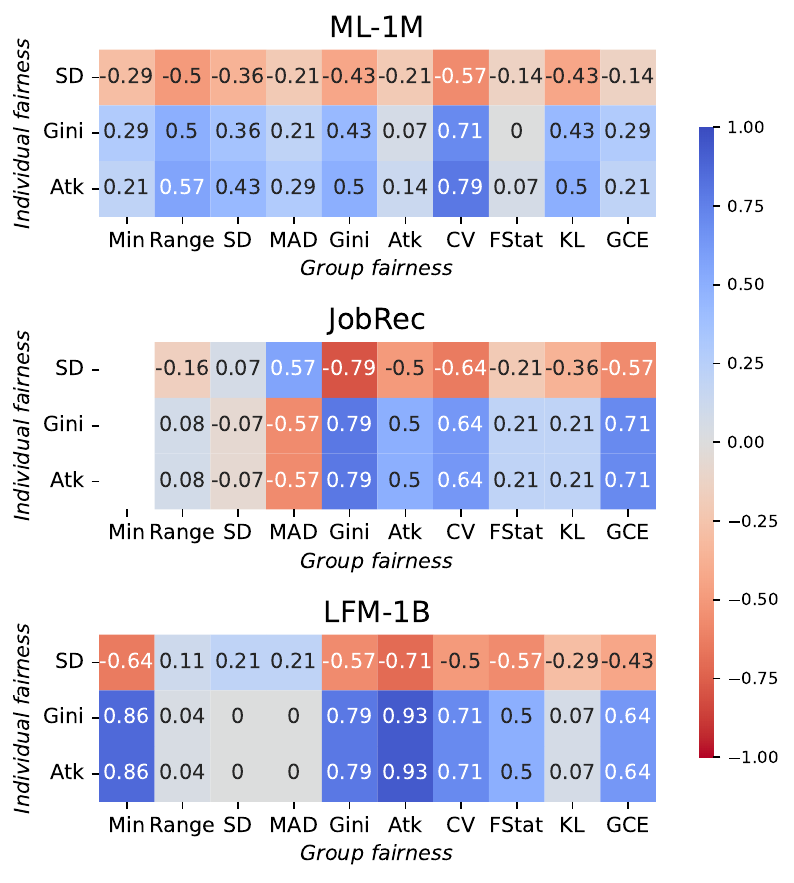}
    \caption{Agreement (Kendall's $\tau$) of NDCG-based measures for individual \fair{} ($y$-axis) and group \fair{} ($x$-axis) in ranking LLMRecs. User groups are based on 3 sensitive attributes. Due to 8-way ties, $\tau$ cannot be computed for Min (JobRec).
    }
    \label{fig:corr_ind_group}
\end{figure}

Do individual and group \fair{} measures reach the same conclusion? 
If an individual \fair{} measure can rank models from the most to the least fair equivalently to a group \fair{} measure, one can be a proxy for the other. 
To assess measure agreement in ranking models, we compute Kendall's $\tau$; we consider two rankings to be ``equivalent'' if $\tau \geq 0.9$ \cite{Maistro2021PrincipledRankings}. 
\Cref{fig:corr_ind_group} shows the agreement of group and individual \fair{} measures. 

Our results show that \textbf{no individual \fair{} measure consistently has equivalent rankings to any group \fair{} measures}.\footnote{We find similar results for the group-individual fairness agreement between the same family of measures (e.g., SD$_\text{b-group}$ vs SD$_\text{ind}$) for all possible groupings (App.~B.2).} For all datasets, group \fair{} measures tend to have weak-to-strong agreement with Gini\ind{}/Atk\ind{} and have weak-to-strong disagreement with SD\ind{}. The only group \fair{} measure that always correlates strongly to an individual \fair{} measure is CV ($\tau \in [0.64,0.79]$). This may be due to the measures' similar formulation: Gini\ind/Atk\ind{} have division by the mean \eff{} scores across all users, while CV has division by the mean of mean \eff{} score per group.
\textbf{In summary, no existing individual \fair{} measures make a reliable proxy for group \fair{} measures, hence the need to evaluate for both.} 

\subsubsection*{Intersectional fairness (RQ2)} 
\begin{figure}[tb]
    \centering
    \includegraphics[width=0.98\linewidth]{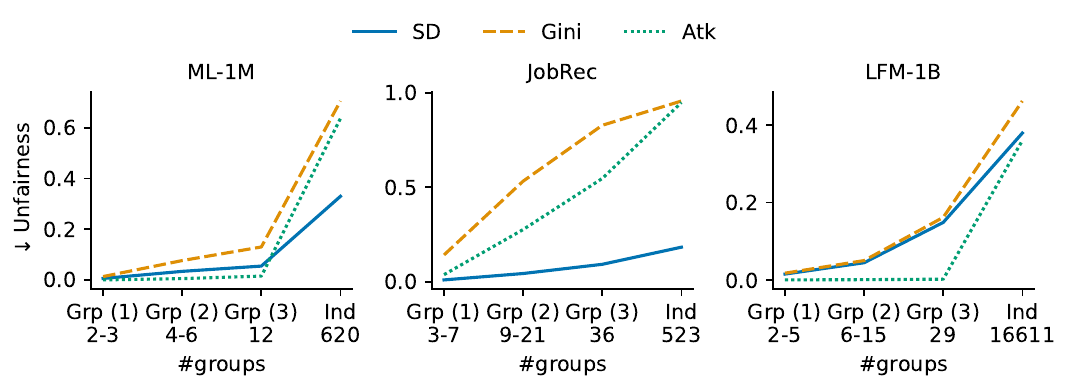}
    \caption{NDCG-based Group (Grp) and individual (Ind) \fair{} scores of GLM-4-9B (NS-prompt). Grp ($a$) is the mean \fair{} scores across all ways of grouping users when considering $a$ attribute(s). For $a\in\{1,2\}$, the \# of ways to group users is three each. For $a=3$, there is only one way to group users.
    }
    \label{fig:diff_groups}
\end{figure}

As we consider more intersectional attributes to form user groups, the number of groups grows. 
We posit that achieving intersectional group fairness is harder than for non-intersectional groups as it requires having similar \eff{} scores for a larger number of groups. Likewise, individual fairness (i.e., having a group for each user) would be harder to achieve than group fairness.  We study: 
(i) if/how the number of attributes used for grouping 
affect fairness; and 
(ii) the difference between individual and (intersectional) group \fair{} scores.  
To this end, we compute NDCG-based individual \fair{} scores across all users, and average NDCG-based group \fair{} scores across all ways of grouping users when considering only 1, 2, or 3 attribute(s) at once. SD, Gini, and Atk are computed, as they have been used as both group and individual \fair{} measures. We evaluate the runs from 
GLM-4-9B (NS) due to their relatively good NDCG and \fair{} scores across all datasets. 
The results are shown in \Cref{fig:diff_groups}.\footnote{While we present the measures in the same plot, their distribution differs, so the figures should not be used to quantify the gap between two families of measures.}

Regarding (i), our results verify that \textbf{fairness worsens as more attributes are used to form groups, highlighting the importance of considering intersectionality in fairness evaluation.} Note that JobRec has higher $\downarrow$Gini/$\downarrow$Atk than ML-1M/LFM-1B as its NDCG is much lower; Gini/Atk accounts for this, but SD does not.

Regarding (ii), we find that \textbf{even when the recommendations are relatively fair for (intersectional) groups, they can be much less fair for individuals} (i.e., for ML-1M and LFM-1B). For all measures and datasets, the individual \fair{} scores are always worse than group \fair{} scores. These findings imply that group \fair{} scores may mask unfairness within groups, even if the within-group variation is considered (e.g., in Atk$_\text{b-group}$).

\begin{figure}
    \centering
    \includegraphics[width=0.98\linewidth, trim=0 2mm 0 0, clip=True]{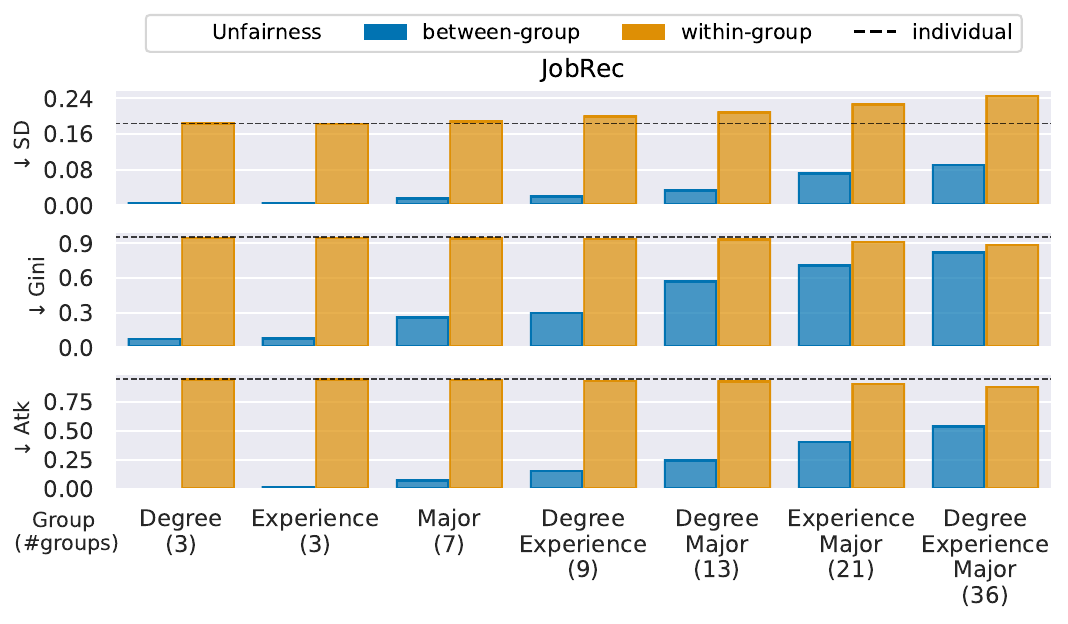}
    \includegraphics[width=0.98\linewidth, trim=0 2mm 0 1cm, clip=True]{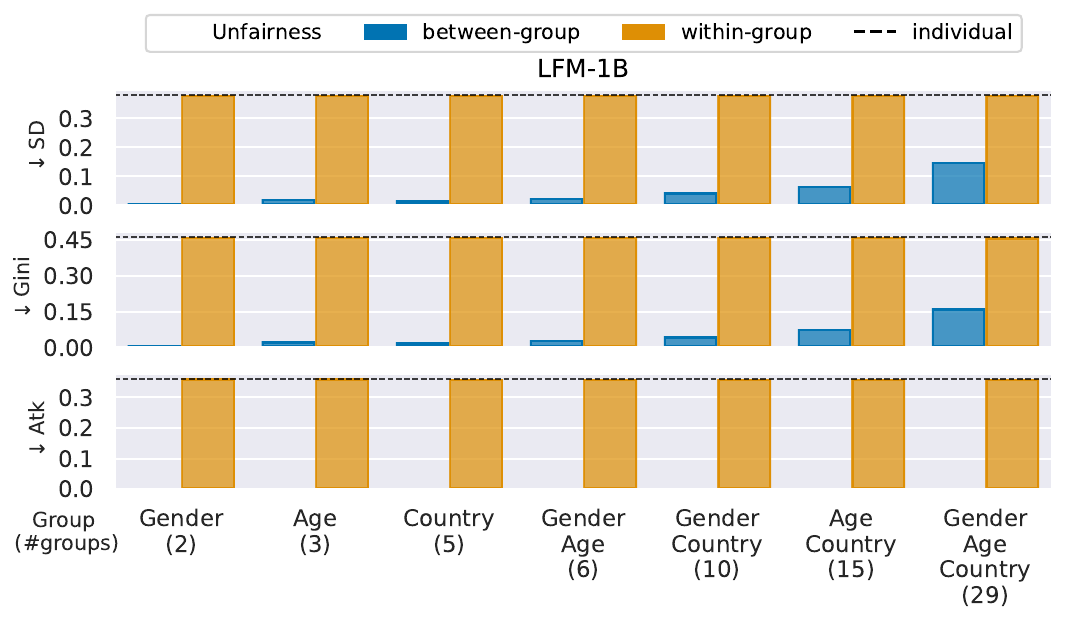}
    \caption{NDCG-based individual, between- and within-group unfairness of GLM-4-9B (NS-prompt) for all ways of grouping users in JobRec and LFM-1B.
    }
    \label{fig:decomposability_jobrec}
\end{figure}

\subsubsection*{Fairness decomposability (RQ3 \& RQ4)}
\label{ss:decompose_analysis}
When fairness is evaluated only between groups, within-group unfairness may occur undetected. As the number of groups increases, we analyse: 
(i) how between- and within- group fairness change; and 
(ii) how they relate to individual fairness. 
To this end, we compute between- and within-group fairness, as well as individual fairness with SD, Gini, and Atk on the NDCG scores from GLM-4-9B (NS).
Results for JobRec and LFM-1B are shown in \Cref{fig:decomposability_jobrec}; we find similar trends for ML-1M (see App.~B.3). 

Regarding (i), we find that while between-group fairness generally worsens as the number of groups grows,\footnote{Resembling stairs; hence, the title of the paper.} 
within-group fairness tends to remain stable, except for SD of JobRec, where it worsens. 
Regarding (ii), we see that within-group unfairness is almost as high as individual unfairness, and always higher than between-group unfairness. Sometimes, within-group unfairness slightly exceeds individual unfairness (e.g., SD for JobRec). In summary, {\bf for different ways of grouping users, within-group fairness is consistently worse than between-group fairness. Yet, individual fairness is generally comparable to within-group fairness, regardless of how the users are grouped.}

\section{Discussion and Conclusion}
RS fairness evaluation often focuses exclusively on group fairness and less on individual fairness or both fairness types \cite{Deldjoo2024FairnessDirections}. We empirically studied the relationship between group and individual fairness in RSs. Our results show that RSs which are fair for groups can still be very unfair for individual users, providing the first empirical evidence on the disjointness of these two RS fairness concepts. Although our experiments are centred on user fairness, we expect similar findings for item (provider) fairness evaluation, considering that measures of user fairness may exhibit similarities to item fairness measures in how scores are aggregated across groups or individuals. Moreover, our findings can be applied directly to real-world RS design and fairness mitigation strategies, for example, by incorporating both between-group and within-group fairness terms in the loss function. 

Overall, we encourage evaluating individual and within-group fairness alongside group fairness, as between-group \fair{} scores may mask huge disparity in RS effectiveness across users, even for a large number of groups (i.e., 7\% of the number of users) and even when the between-group \fair{} score accounts for within-group variations. Future work will study how strategies that mitigate either group or individual fairness affect each fairness type.

\begin{acks}
This work is supported by the Algorithms, Data, and Democracy project (ADD-project), funded by the Villum Foundation and Velux Foundation. This work is also supported in part by the Australian Research Council (DP190101113). 
We thank the anonymous reviewers who have provided helpful feedback to improve an earlier version of the manuscript.
\end{acks}

\bibliographystyle{ACM-Reference-Format}
\balance
\bibliography{references}

\end{document}